\documentclass[a4paper,10pt,twoside]{cpc-hepnp}
\usepackage{multicol}
\usepackage{graphicx}
\usepackage{booktabs}
\usepackage{amssymb,bm,mathrsfs,bbm,amscd}
\usepackage[tbtags]{amsmath}
\usepackage{lastpage}

\usepackage{overpic} 
\usepackage{xcolor}

\begin{document}


\fancyhead[c]{\small Submitted to `Chinese Physics C' }
\fancyfoot[C]{\small 010201-\thepage}


\title{Development of a portable and fast wire tension measurement system for MWPC's construction\thanks{Supported by National Natural Science Foundation of China (A050506), State Key Laboratory of Particle Detection and Electronics and Key Laboratory of China Academy of Engineering Physics (Y490KF40HD) }}

\author{%
      Jing-Hui Pan$^{1,2,3}$
\quad Chang-Li Ma$^{2,3;1)}$\email{machangli@ihep.ac.cn}%
\quad Xue-Yu Gong$^{1;2)}$\email{gongxueyu@126.com}%
\quad Zhi-Jia Sun$^{2,3,4;3)}$\email{sunzj@ihep.ac.cn}\\%
\quad Yan-Feng Wang$^{2,3}$
\quad Chen-Yan Yin$^{1}$
\quad Lei Gong$^{5}$
}
\maketitle

\address{%
$^1$ University of South China£¬Hengyang 421001, China\\
$^2$ Institute of High Energy Physics, Chinese Academy of Sciences, Beijing 100049, China\\
$^3$ Dongguan Neutron Science Center, Dongguan 523803, China\\
$^4$ State Key Laboratory of Particle Detection and Electronics, Beijing 100049, China\\
$^5$ Fujian Fuqing Nuclear Power Co., Ltd, China National Nuclear Corporation, Fuqing 350318, China
}

\begin{abstract}
In a multi-wire proportional chamber detector(MWPC), the anode and signal wires must maintain suitable tensions, which is very important for the detector's stable and perfect performance. As a result, wire tension control and measurement is essential in MWPC's construction. The thermal neutron detector of multi-functional reflectometer at China Spallation Neutron Source is designed using a high pressure $^{3}$He MWPC detector, and in the construction of the detector, we developed a wire tension measurement system. This system is accurate, portable and time-saving. With it, the wires' tension on a anode wire plane has been tested, the measurement results show that the wire tension control techniques used in detector manufacture is reliable.
\end{abstract}



\footnotetext[0]{\hspace*{-3mm}\raisebox{0.3ex}{$\scriptstyle\copyright$}2013
Chinese Physical Society and the Institute of High Energy Physics
of the Chinese Academy of Sciences and the Institute
of Modern Physics of the Chinese Academy of Sciences and IOP Publishing Ltd}%

\begin{multicols}{2}

\section{Introduction}

China Spallation Neutron Source(CSNS) is under construction, and three neutron scattering instruments are built at the same time, belong which, the multi-functional reflectometer employs a high pressure $^3$He gas multi-wire proportional chamber(MWPC) as its neutron detector\cite{lab1}. This detector is designed using an 8 atm $^3$He/C$_3$H$_8$(80/20) mixture as working gas and gold-plated tungsten as anode and signal wires. The sensitive area of the detector is designed to be 200mm$\times$200mm, and the neutron space resolution is expected less than 2mm, besides the counting rate can reach to 10$^7$/cm$^2$s\cite{lab2}. As the influence of electrostatic force and gravity, the anode and signal wires will have position offset, as a result the performance of the detector, including magnification, position resolution, etc, will be affected\cite{lab3}. To ensure the wires' position deviation as small as possible£¬it is very important to keep the wires with proper tensions, so we have to measure and control the wire tension in the detector construction. The principle of the wire tension measurement is based on the relationship between the tension and its vibration inherent frequency, which is shown as formula (\ref{eq1}):
\begin{eqnarray}
\label{eq1}
T=1\times10^{-6}\cdot\rho(2Lf_{0})^{2}.
\end{eqnarray}

Where $T$(N) is the wire tension, $f_{0}$(Hz) is the wire vibration fundamental frequency, $\rho$(mg/m) is the linear mass density, and $L$(m)is the wire length.

Based on the principle, a lot of equipment was designed, and in general they can be sorted into two classes. In the first class, it inputs different periodic driving force to a wire, and make the wire resonance, and then by determining the resonance to measure the wire inherent frequency. Because of driving force generator and resonance determination module, these equipment is complicated and the measurement is time consuming\cite{lab4,lab5,lab6,lab7,lab8,lab9,lab10}. In the other class, a short-time driving force was used to make the wire vibrate, and then measuring the signal of the vibration to analysis the wire inherent frequency\cite{lab11,lab12,lab13,lab14,lab15}. In these equipment, the wire inherent frequency is measured directly, so the measurement is quicker and more accurate, but they need a powerful signal processing system.

In this paper, we present a simple, quick and accurate equipment by adopting the wires' vibration method for measuring the wires' tension in MWPC construction.

\section{Description of tension measurement system}

According to Faraday's law£¬when a wire with current (I) is placed in a magnetic field (B), it will experience a force (F), and the force is equal to I$\cdot L \times$B. Meanwhile, when a wire is vibrating in a magnetic field with velocity $\nu$, electric potential is induced along the wire. The potential is equal to $L\cdot \nu\times$B. Based on this principle, we designed a simply equipped wire tension measurement system: two electromagnets are used to generate an adjustable magnetic field, and the wire which will be measured is placed between them. A mono-stable trigger mode is used to produce narrow pulses, then the pulses are amplified by a power
amplifier and used to stimulate the wire to vibrate in the magnetic field. To make sure the wire generates largest vibration, the pules' widths are controlled to about one fourth of the expected wire inherent vibration period. Then the vibration signal, which is an attenuate sine wave in theory, is amplified by a operational amplifier module. At last, the vibration signal is recorded and analyzed by a digital oscilloscope or a computer controlled data acquisition board, and the vibration frequency is measured. A schematic drawing of the system is shown in Fig.\ref{fig1}.

\begin{center}
\includegraphics[width=0.47\textwidth]{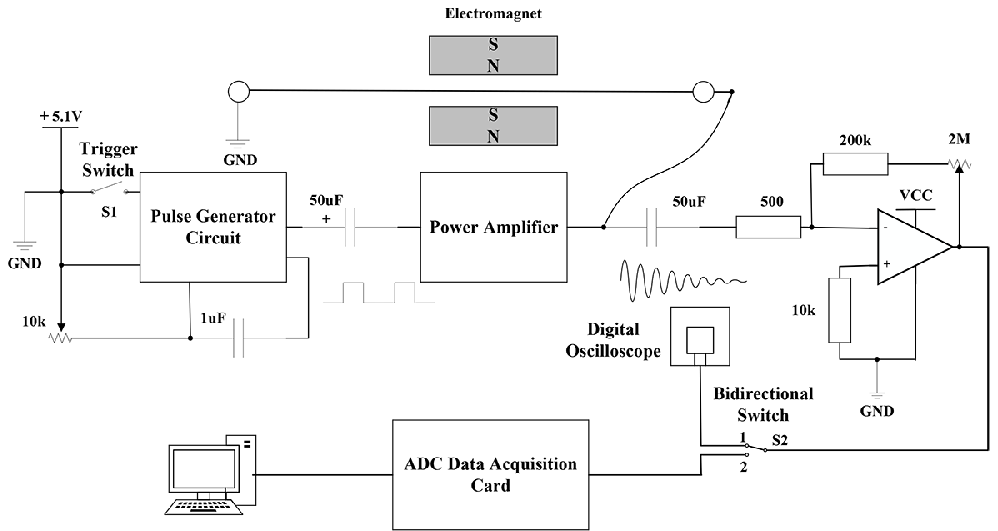}
\figcaption{\label{fig1} Schematic block diagram of the wire tension measurement system. }
\end{center}

As shown in Fig.\ref{fig1}, two solenoidal electromagnets with 22mm diameter magnetic poles, which are operated with a 30V DC power supply, are placed in the same direction with about 10mm distance. When the power supply's voltage is set to 30V, the magnetic field  between the magnets along the wire's direction is shown in Fig. \ref{fig2}. If the current in a wire is 1A, the force applied to the wire is about $3.24\times10^{-3}$N. The mono-stable trigger, which is connected with a 1$\mu$F capacitor and a 10K$\Omega$ variable resistor and operated with a 5V DC power, can generate TTL pulse with a maximum width of 4ms, which is calculated by the formula, $\tau = K \cdot R\cdot C$, ($K=0.4$, which is from
mono-stable trigger chip product manual). It makes the system adapt to measure wires with inherent vibration frequency no less than 62.5Hz. The power amplifier in the system has a 3A rated current and can amplify the a TTL pulse to 15V level. In the operational amplifier module, a 500$\Omega$ resistor is connected at input terminal and a 2M$\Omega$ potentiometer is used as feedback resistor, so the voltage amplification factor can get up to 4000. For flexibility, we design two options to analyze vibration signals. One method is using a digital oscilloscope to record the signal and do Fourier transform, and the other alternative is using a 12 bits ADC data acquisition board to converts the analog signals to digital signals and transmits them to computers, then a LabVIEW program is used to display and analyze the signals. To suppress direct current interference, 50$\mu$F capacitors are used among different function modules.

\begin{center}
\includegraphics[width=0.47\textwidth]{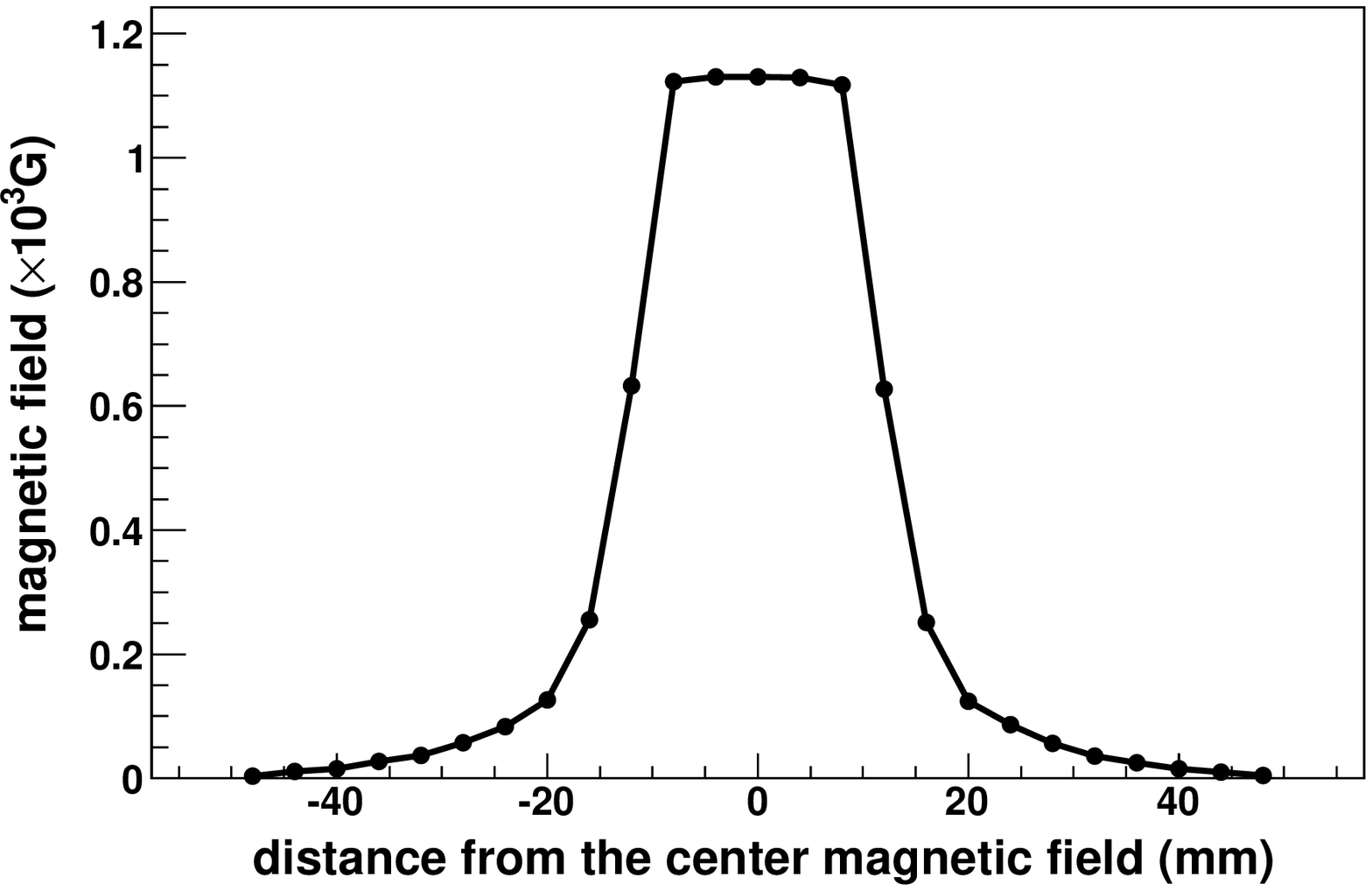}
\figcaption{\label{fig2} The magnetic field distribution between two magnets along the measured wire. }
\end{center}

\begin{center}
\setlength{\abovecaptionskip}{0.pt}
\setlength{\belowcaptionskip}{-0.pt}
\includegraphics[width=0.47\textwidth]{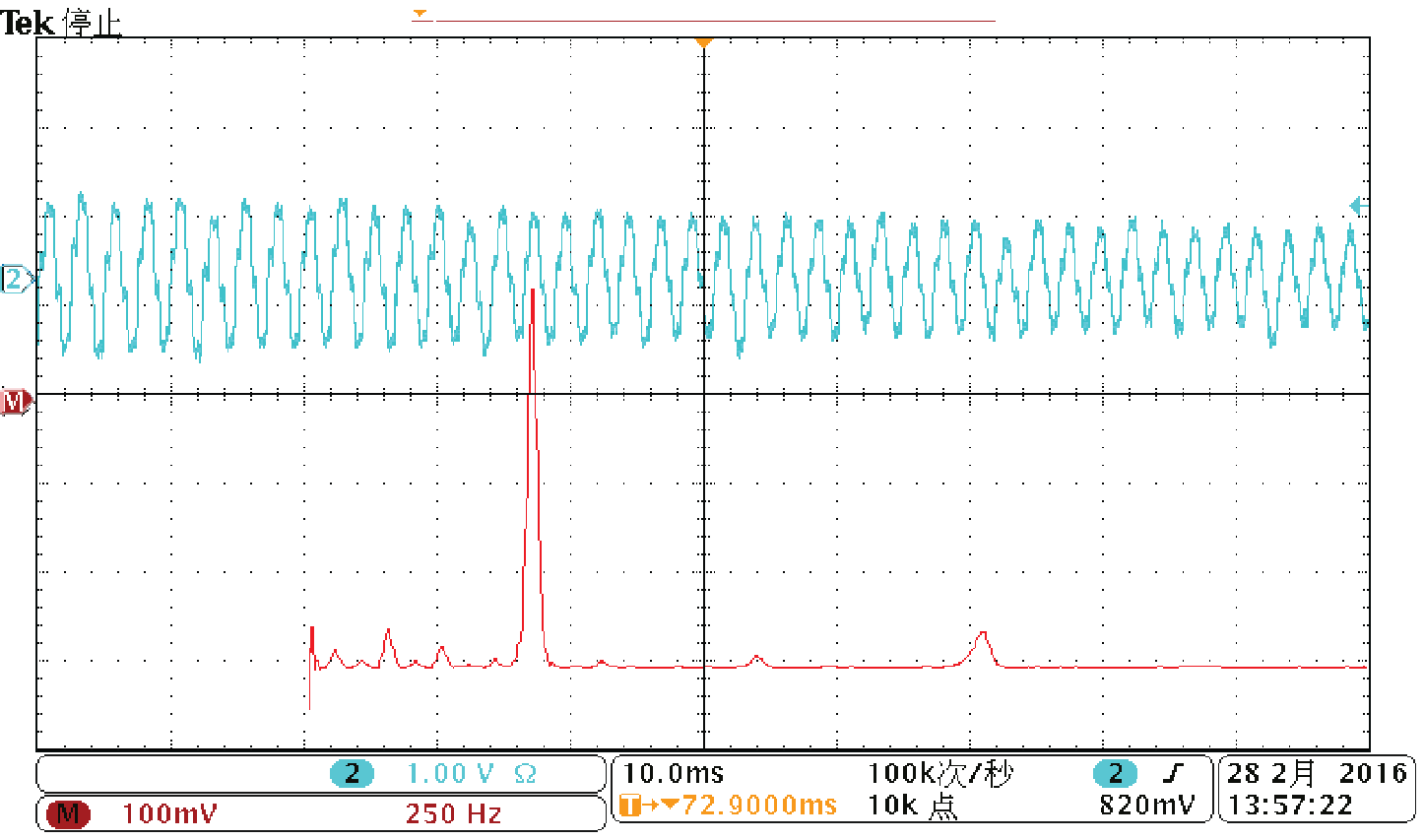}
\figcaption{\label{fig3} The upper waveform is a wire's vibration signal and the lower one is the corresponding spectrum of the Fourier transform which is  analyzed by a digital oscilloscope. }
\end{center}

A vibration signal is captured by oscilloscope and its frequency spectrum are shown
in Fig.\ref{fig3}. In this picture, the wire's diameter, length and applied tension
are 25$\mu$m, 240mm and 40g respectively, and the operation voltage of the
electromagnets is set at 30V. The highest peak at about 420Hz on the frequency spectrum is the fundamental harmonic, while the peaks at about 1260Hz is the third harmonic of the wire. As this, measuring a wire's frequency with the system takes only a few seconds. At the same time, this system records a half first harmonic signal\cite{lab10}.

For flexibility, an ADC data acquisition board and a LabVIEW computer program
are designed as an alternative to a digital oscilloscope. The data acquisition board has a sample rate of 80KHz and can convert $\pm$8V analog signals into digital ones. With a USB2.0 connector digital signals in the board are transmitted to a personal computer. Using a LabVIEW signal analysis program, the measurement digital accuracy can reach to 1Hz, which is on the same order of the measurement error. Only taking a low-voltage power supply and a personal computer, the system can be used at any
places we need. Fig.\ref{fig4} are the front panel and the flow diagram of its corresponding graphical program in the LabVIEW platform respectively.

\end{multicols}
\begin{center}
\setlength{\abovecaptionskip}{0.pt}
\setlength{\belowcaptionskip}{-0.pt}
\includegraphics[width=8.2cm,height=6.7cm]{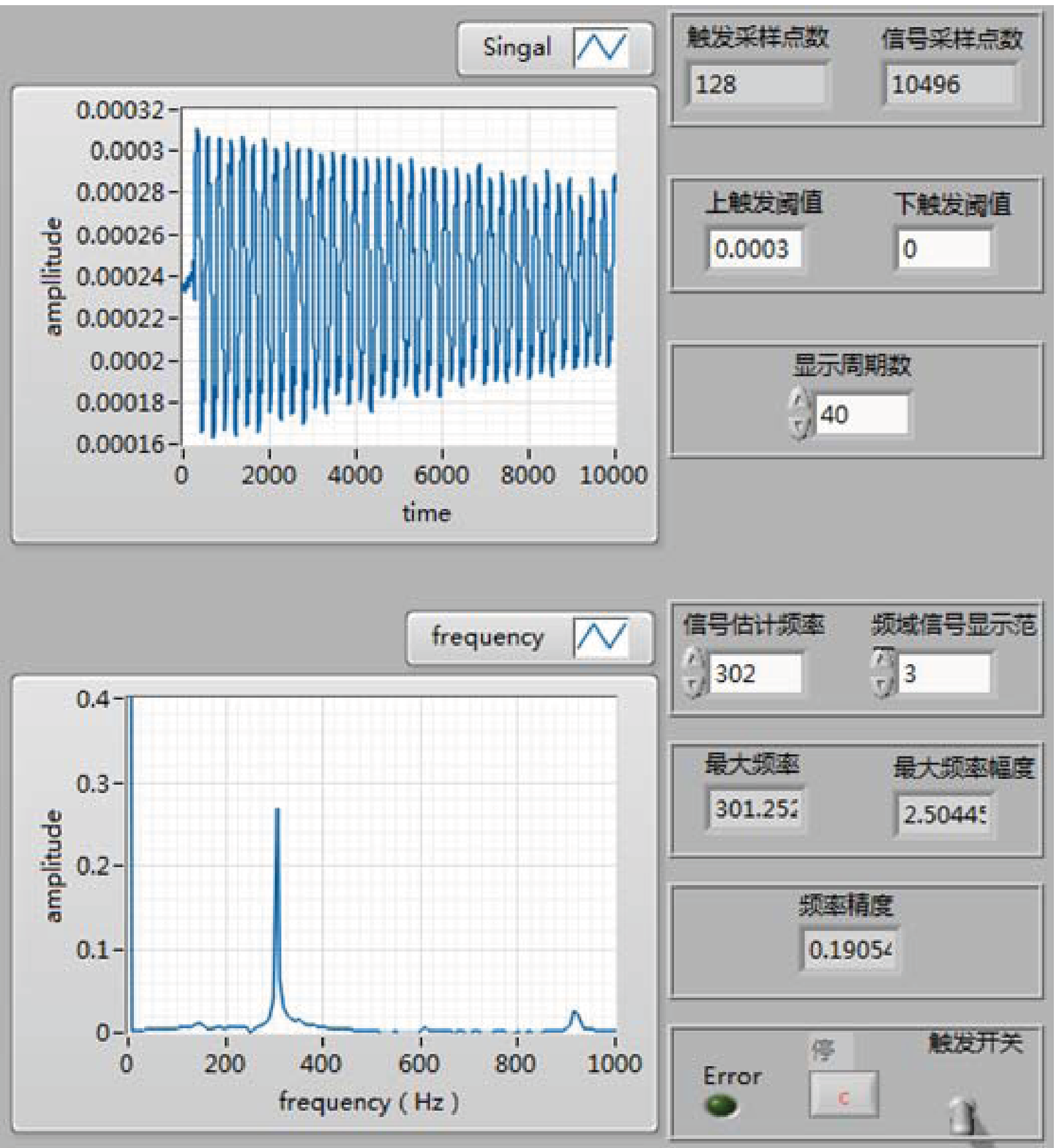}
\includegraphics[width=8.2cm,height=6.7cm]{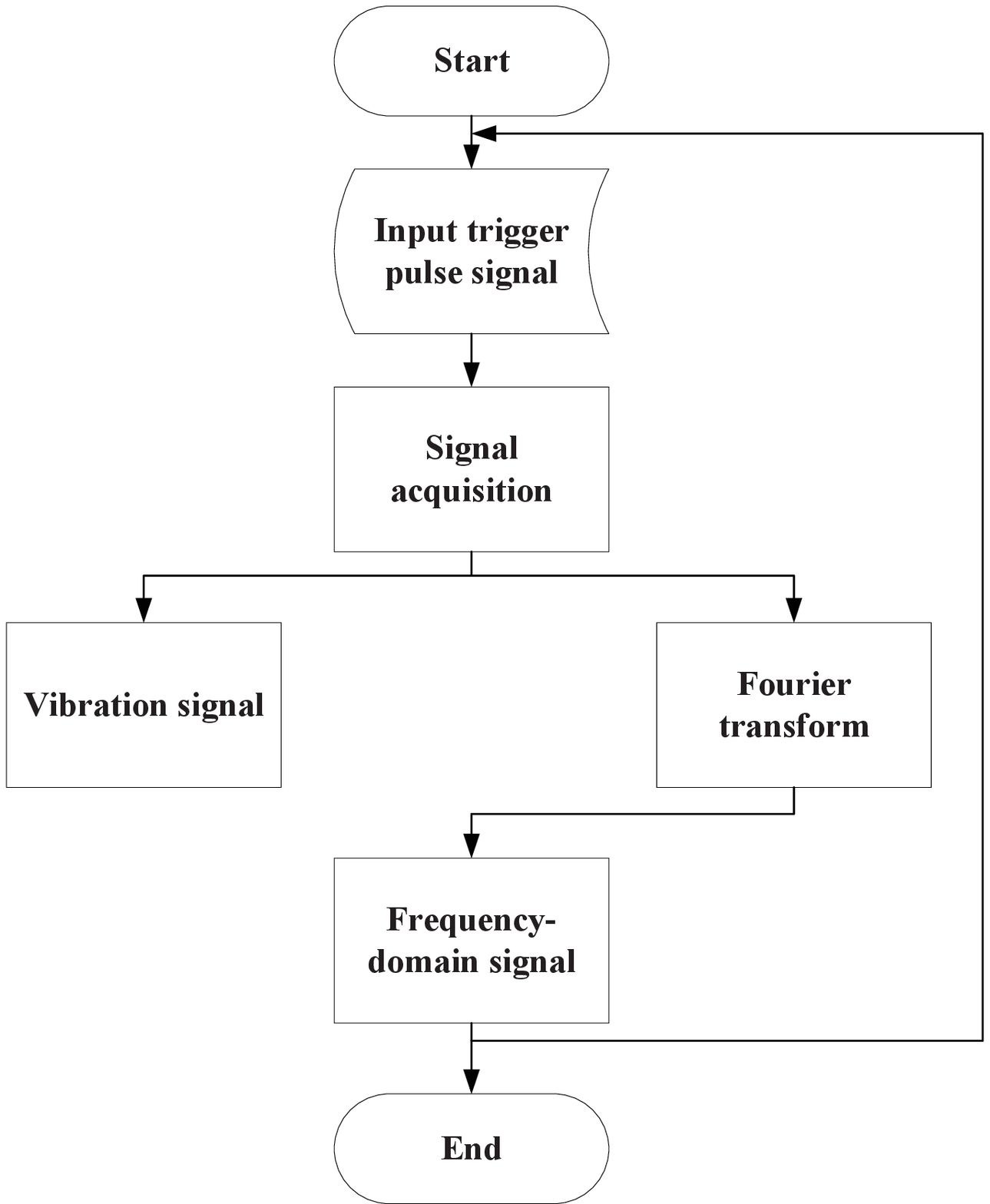}
\figcaption{\label{fig4} The LabVIEW program of the wire tension  measurement system (left: front panel, right: flow diagram of the graphical program).}
\end{center}
\begin{multicols}{2}

\begin{center}
\setlength{\abovecaptionskip}{0.pt}
\setlength{\belowcaptionskip}{-0.pt}
\includegraphics[width=0.47\textwidth]{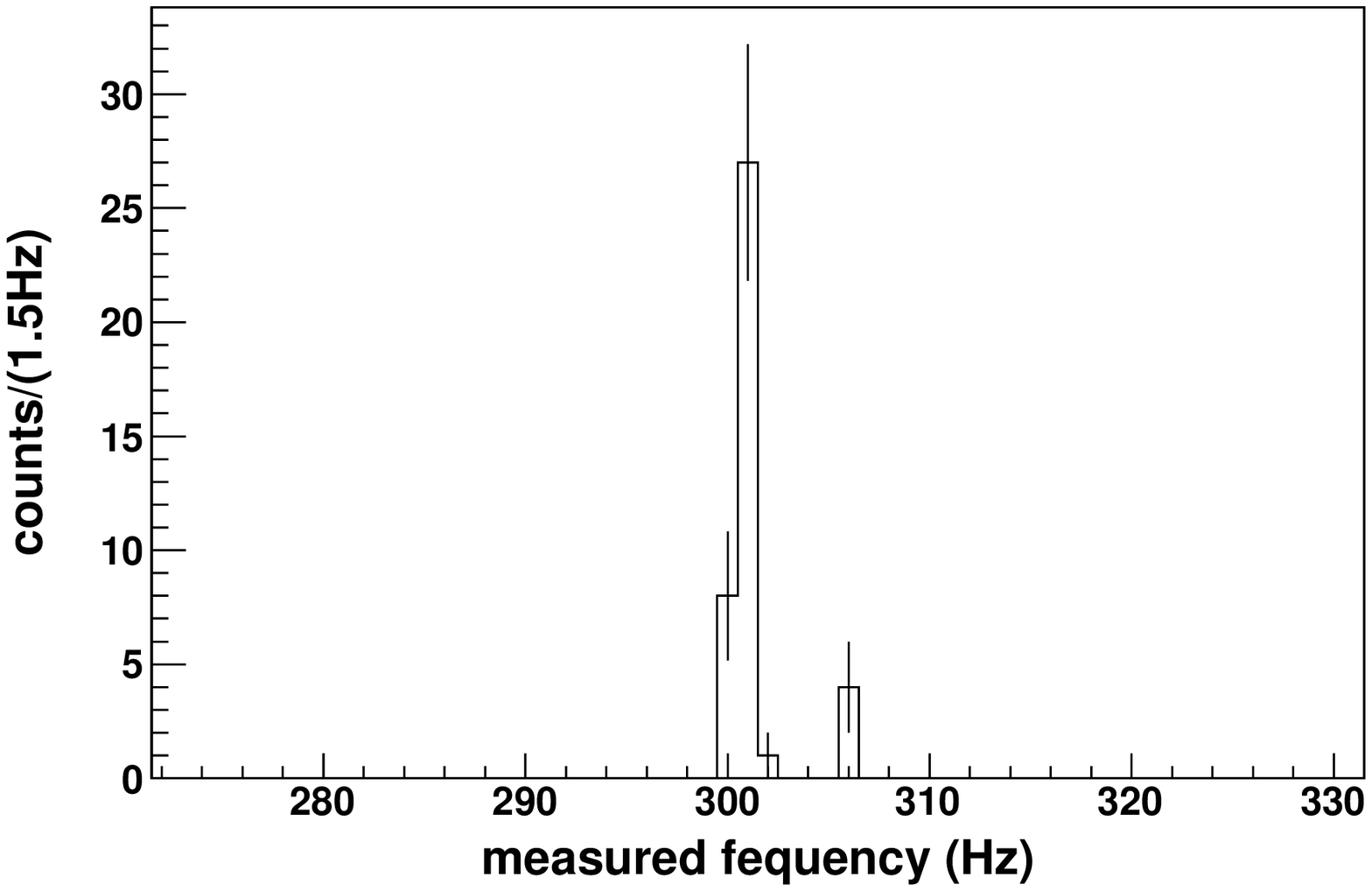}
\figcaption{\label{fig5} The distribution of repeated measurements of a single wire with 20g tension.}
\end{center}

\begin{center}
\setlength{\abovecaptionskip}{0.pt}
\setlength{\belowcaptionskip}{-0.pt}
\includegraphics[width=0.47\textwidth]{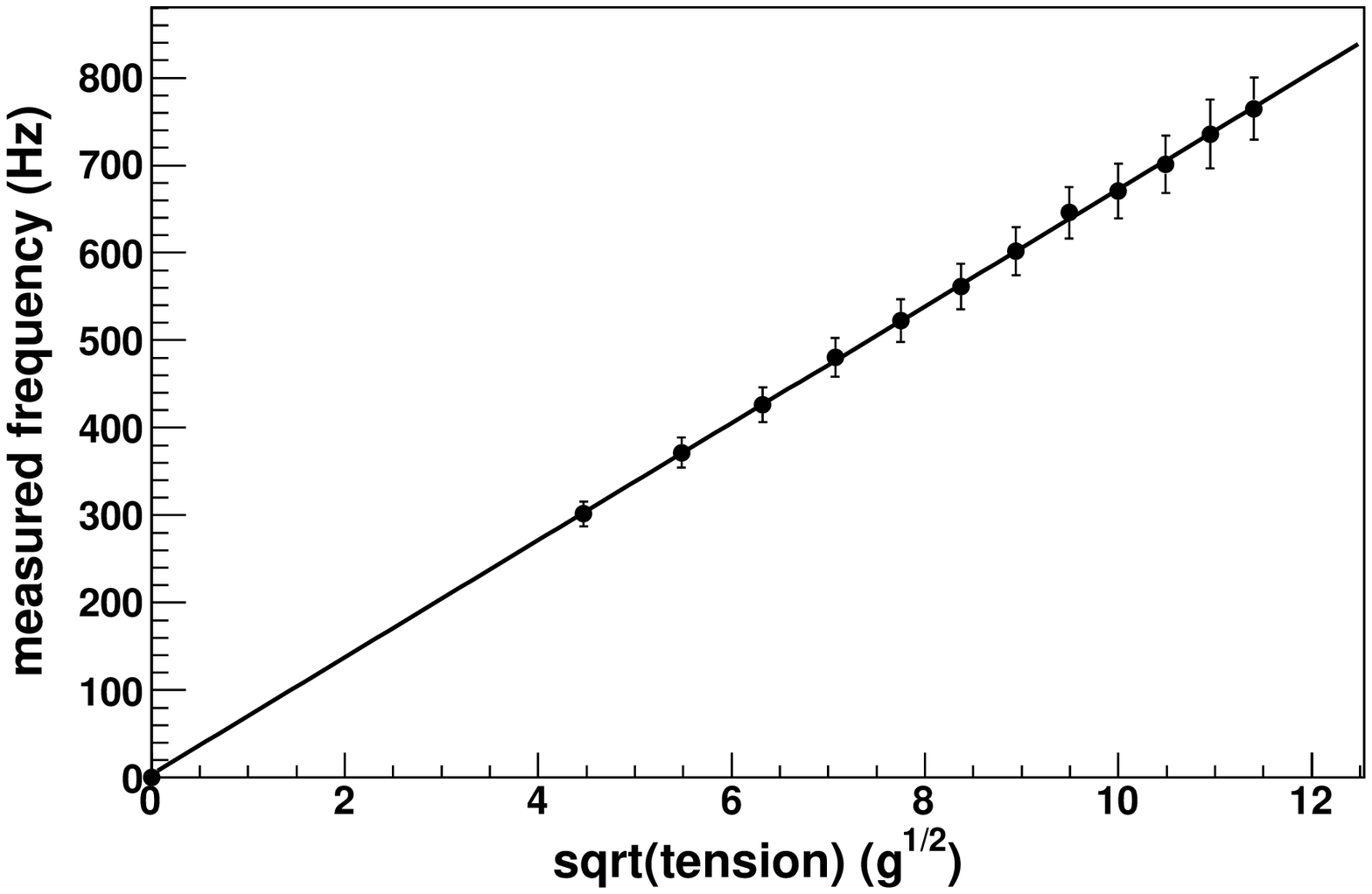}
\figcaption{\label{fig6} The measured frequencies v.s. the wire tensions. X axis is the wire tension's square root and Y axis is the measured wire vibration frequencies.}
\end{center}

\end{multicols}
\begin{center}
\setlength{\abovecaptionskip}{0.pt}
\setlength{\belowcaptionskip}{-0.pt}
\tabcaption{\label{tab1.}The calculated  and measured frequencies and their errors of some wires.}
\footnotesize
\begin{tabular}{c | c c c c c c c c c c c c }
\toprule
\multicolumn{1}{c|}{Tension(g)} &
\multicolumn{1}{c}{20} &
\multicolumn{1}{c}{30} &
\multicolumn{1}{c}{40} &
\multicolumn{1}{c}{50} &
\multicolumn{1}{c}{60} &
\multicolumn{1}{c}{70} \\
\hline
calculated frequency(Hz)& 302$\pm$14& 370$\pm$17& 428$\pm$20& 478$\pm$22& 524$\pm$24& 566$\pm$26\\
measured frequency(Hz)& 301.5$\pm$1.7& 371.4$\pm$1.3& 426.3$\pm$1.1& 480.5$\pm$0.6& 522.4$\pm$2.5& 561.5$\pm$1.2\\
\hline
Tension(g)& 80& 90& 100& 110& 120& 130\\
\hline
calculated frequency(Hz)& 605$\pm$28& 642$\pm$30& 676$\pm$31& 709$\pm$33& 741$\pm$34& 771$\pm$36\\
measured frequency(Hz)& 601.8$\pm$1.3& 645.9$\pm$1.2& 670.5$\pm$1.7& 700.8$\pm$0.8& 735.6$\pm$1.2& 764.9$\pm$1.8\\
\bottomrule
\end{tabular}
\end{center}
\begin{multicols}{2}

\section{Performance of the measurement system}
\subsection{Accuracy  of the system}

For testing the accuracy of the wire tension measurement system,
we measured 240mm long and 25$\mu$m diameter wires with 20, 30, 40\dots 
130g tension 40 times respectively. The linear mass density is $\rho $=9.3$\pm$0.8mg/m, which is provided by the wire's supplier. Fig.\ref{fig5} is the measured frequency results of a 20g tension wire, and the measured frequencies of wires with other tension values have similar distribution. The measured average frequencies and their standard deviations are listed in Tab.\ref{tab1.}, which are marked as ``measured frequency''. In the measurement, the electromagnet voltage was set to 25V. Based on formula (1), we calculated the inherent frequency of every wire, and the results are shown in Tab.\ref{tab1.} too. In Tab.\ref{tab1.}, the calculation error is from linear density uncertainty of the wires. As shown in Tab.\ref{tab1.}, the measured results are very close to the calculated values. We draw the results in Fig.\ref{fig6}  and fit the results with a line. As shown in the picture, the measurement system can distribute different wire tensions perfectly.

\subsection{Signal-to-noise ratio V.S. magnetic intensity}

\begin{figure}[t!]
\begin{minipage}[t]{0.48\linewidth}
\begin{center}
\begin{overpic} [height=0.23\textheight,width=0.9\linewidth]
{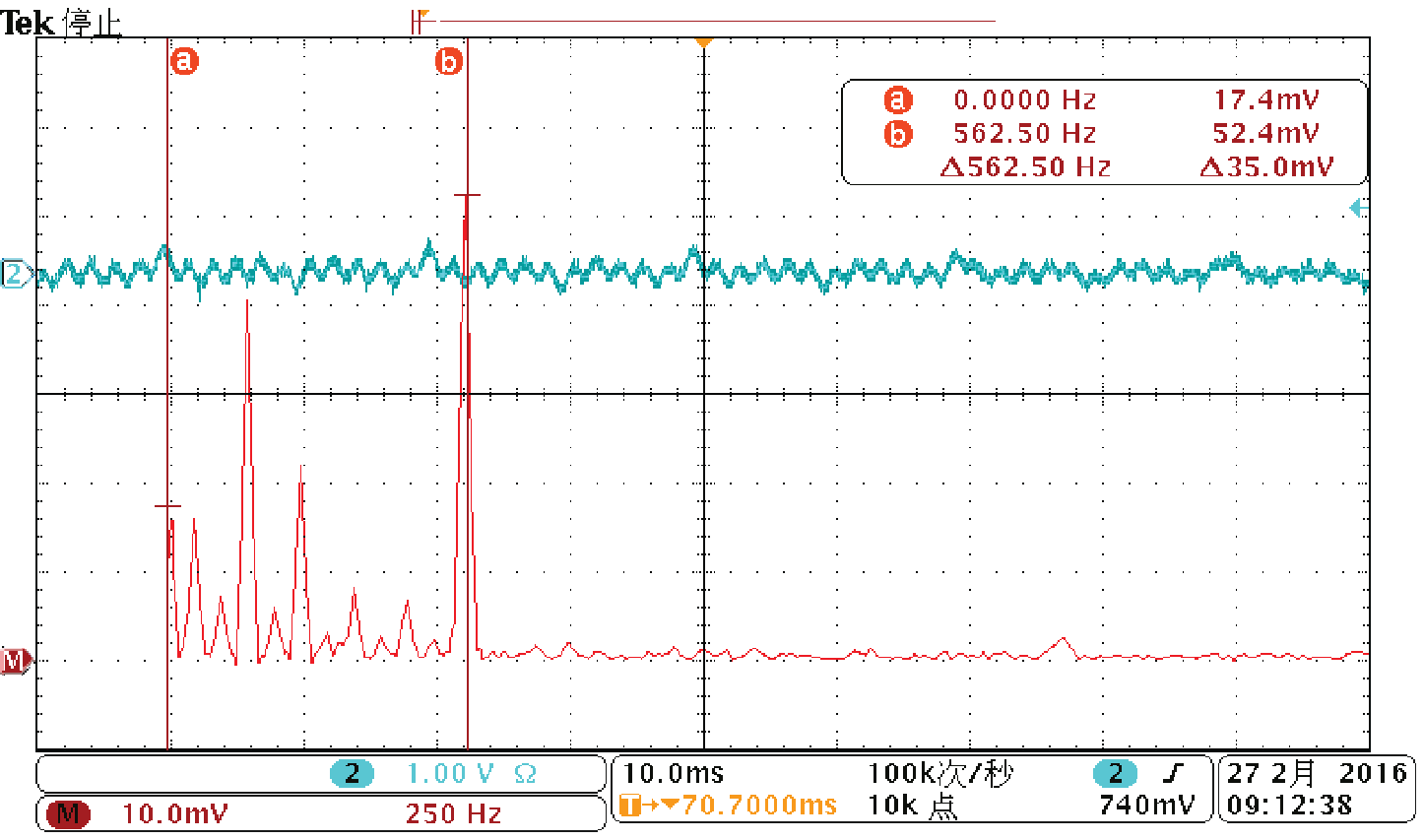}
\put(4.5, 55){(a)}
\end{overpic}
\end{center}
\end{minipage}
\begin{minipage}[t]{0.48\linewidth}
\begin{center}
\begin{overpic} [height=0.23\textheight,width=0.9\linewidth]
{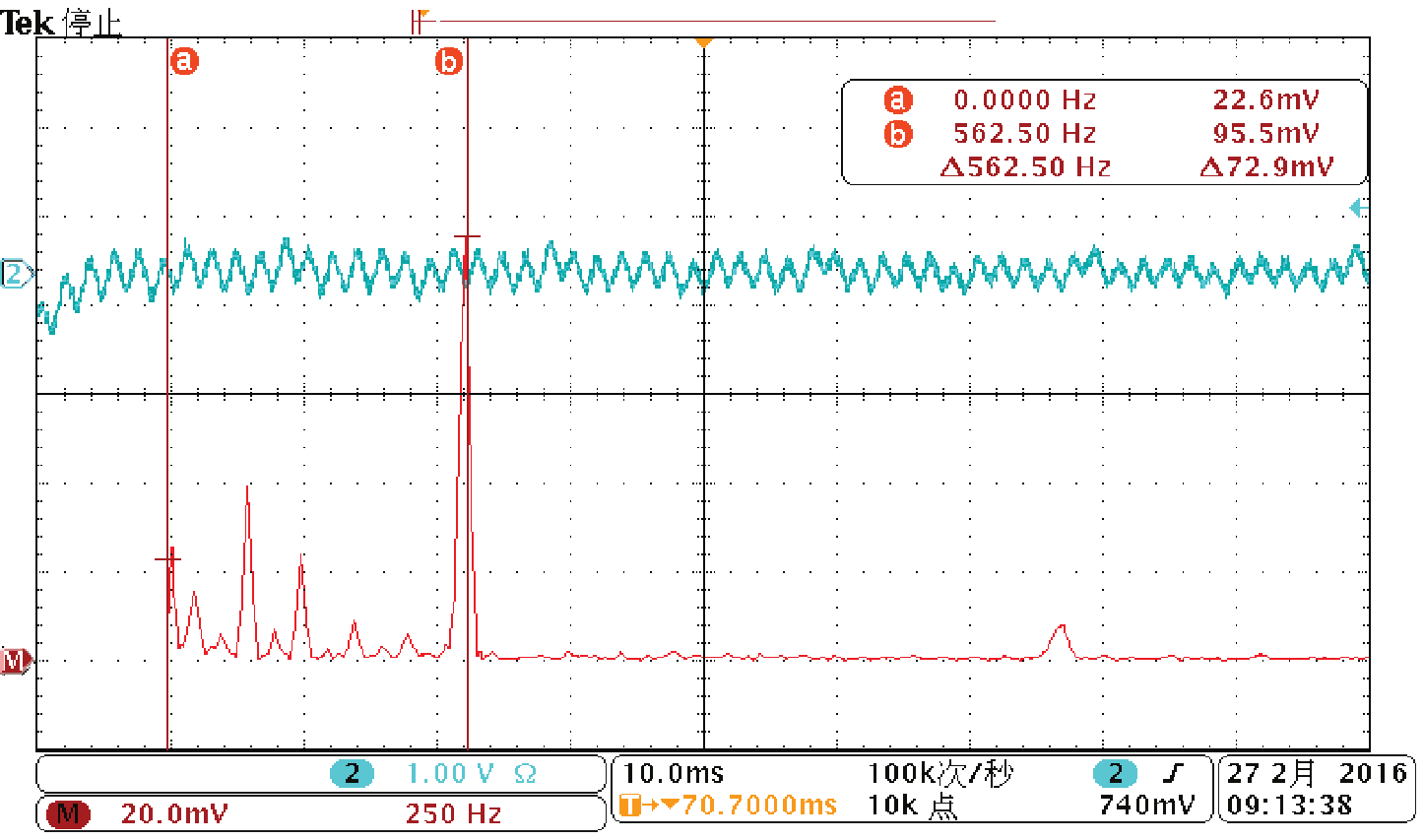}
\put(4.5, 55){(b)}
\end{overpic}
\end{center}
\end{minipage}

\begin{minipage}[t]{0.48\linewidth}
\begin{center}
\begin{overpic} [height=0.23\textheight,width=0.9\linewidth]
{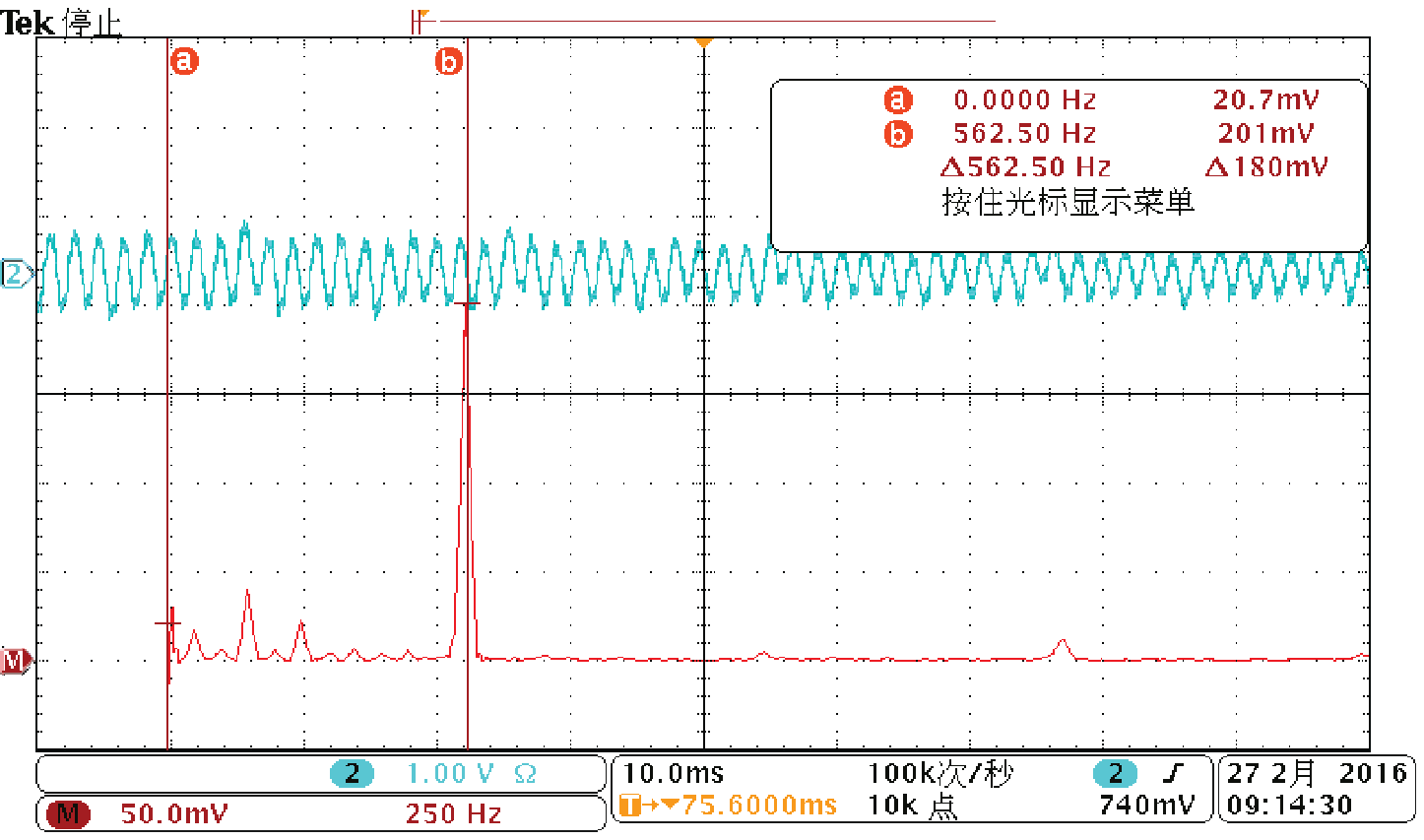}
\put(4.5, 55){(c)}
\end{overpic}
\end{center}
\end{minipage}
\begin{minipage}[t]{0.48\linewidth}
\begin{center}
\begin{overpic} [height=0.23\textheight,width=0.9\linewidth]
{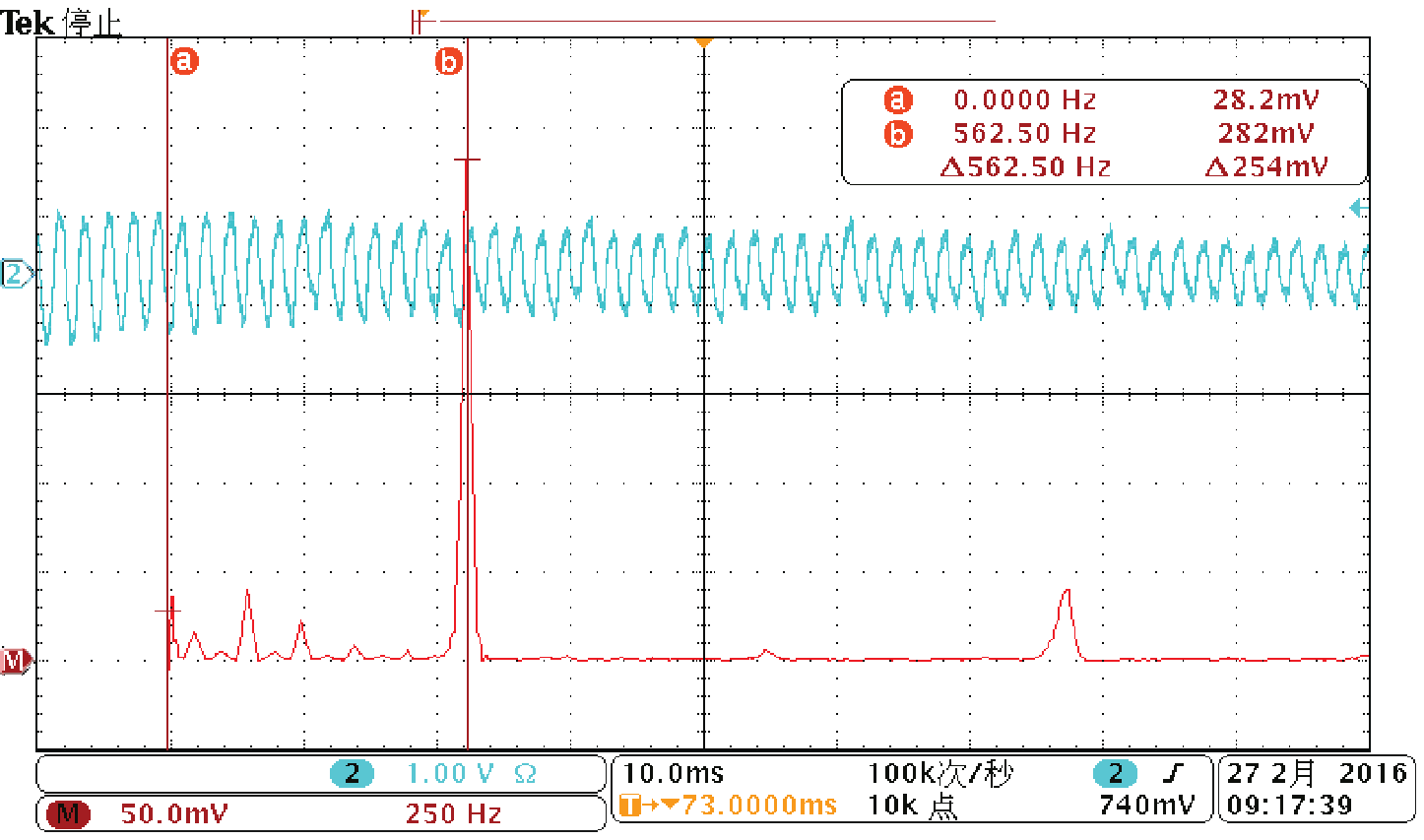}
\put(4.5, 55){(d)}
\end{overpic}
\end{center}
\end{minipage}

\caption{
  \label{fig7}(a),(b),(c),(d) are vibration signals and corresponding frequency spectra when the electromagnets' voltage is set at 10, 15, 20, and 30V respectively.}
\end{figure}

As shown in Ref\cite{lab11}, by enhancing  magnetic field intensity, we can increase the amplitudes of wires'
vibration signals and signal to noise ratio. So we set the electromagnets' voltage
at different values for checking the system performance. Fig.\ref{fig7} (a)-(d) are the vibration signals and their frequency waveforms of wire with 25um diameter, 240mm long and 80g tension when the electromagnet's voltage is set at 10, 15, 20, 30V respectively. As expected, the signal amplitude and the signal to noise ratio increased obviously with the magnetic field's enhancement. Fig.\ref{fig8} is shown the signal amplitudes and measured frequencies with different magnetic intensities. Fig.\ref{fig9} shows that the measurement errors. Obviously, for getting better results, the system should set the magnetic field as strong as possible.

The effect of trigger pulse width on the wire vibration signal is tested too, as expected, it is similar to that of magnetic intensity. The dependence of the vibration signal amplitude on the trigger pulse amplitude are shown in Fig.\ref{fig10}.

\end{multicols}
\begin{multicols}{2}

\end{multicols}
\begin{center}
\setlength{\abovecaptionskip}{0.pt}
\setlength{\belowcaptionskip}{-0.pt}
\includegraphics[width=0.47\textwidth]{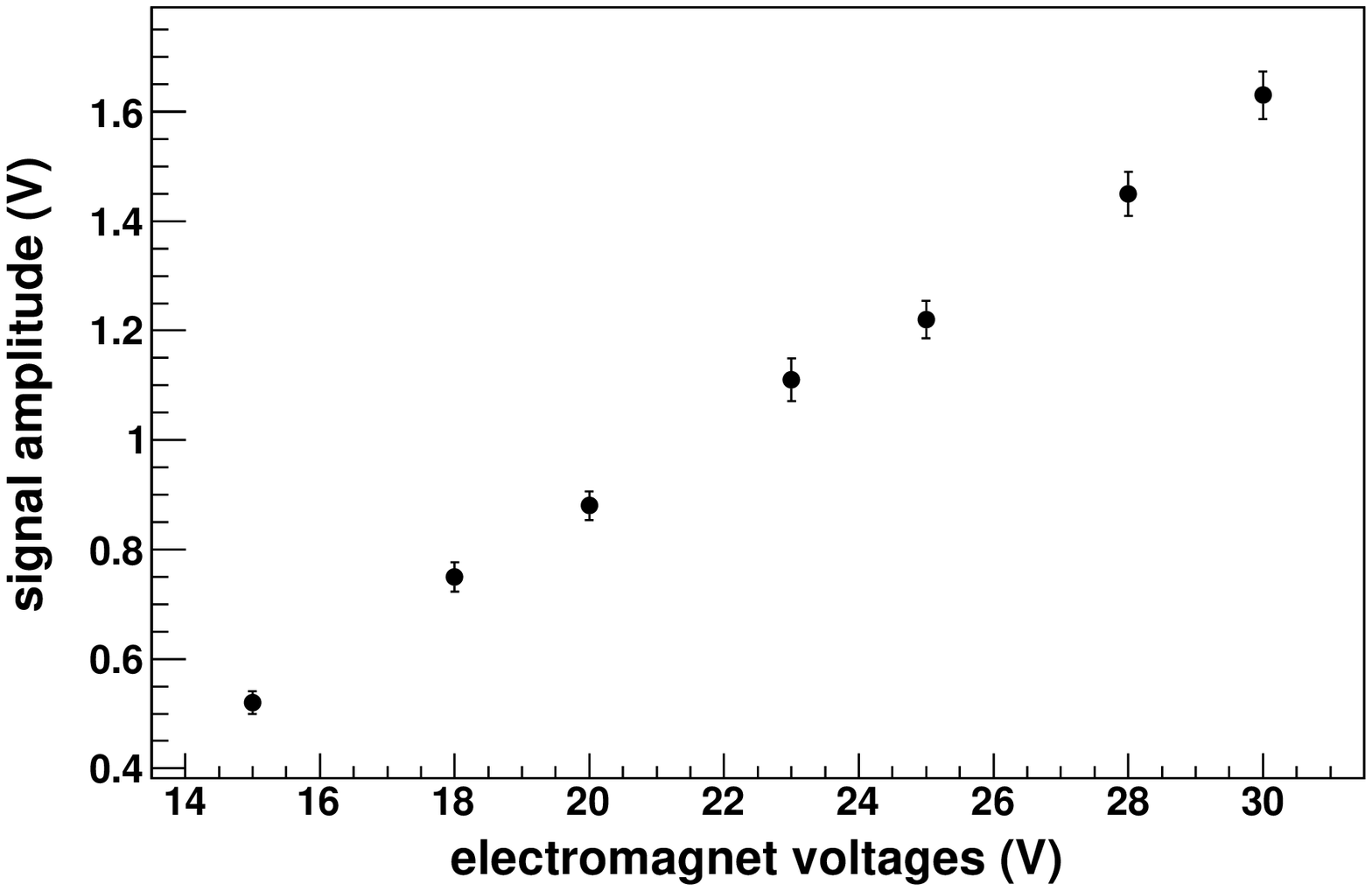}
\includegraphics[width=0.47\textwidth]{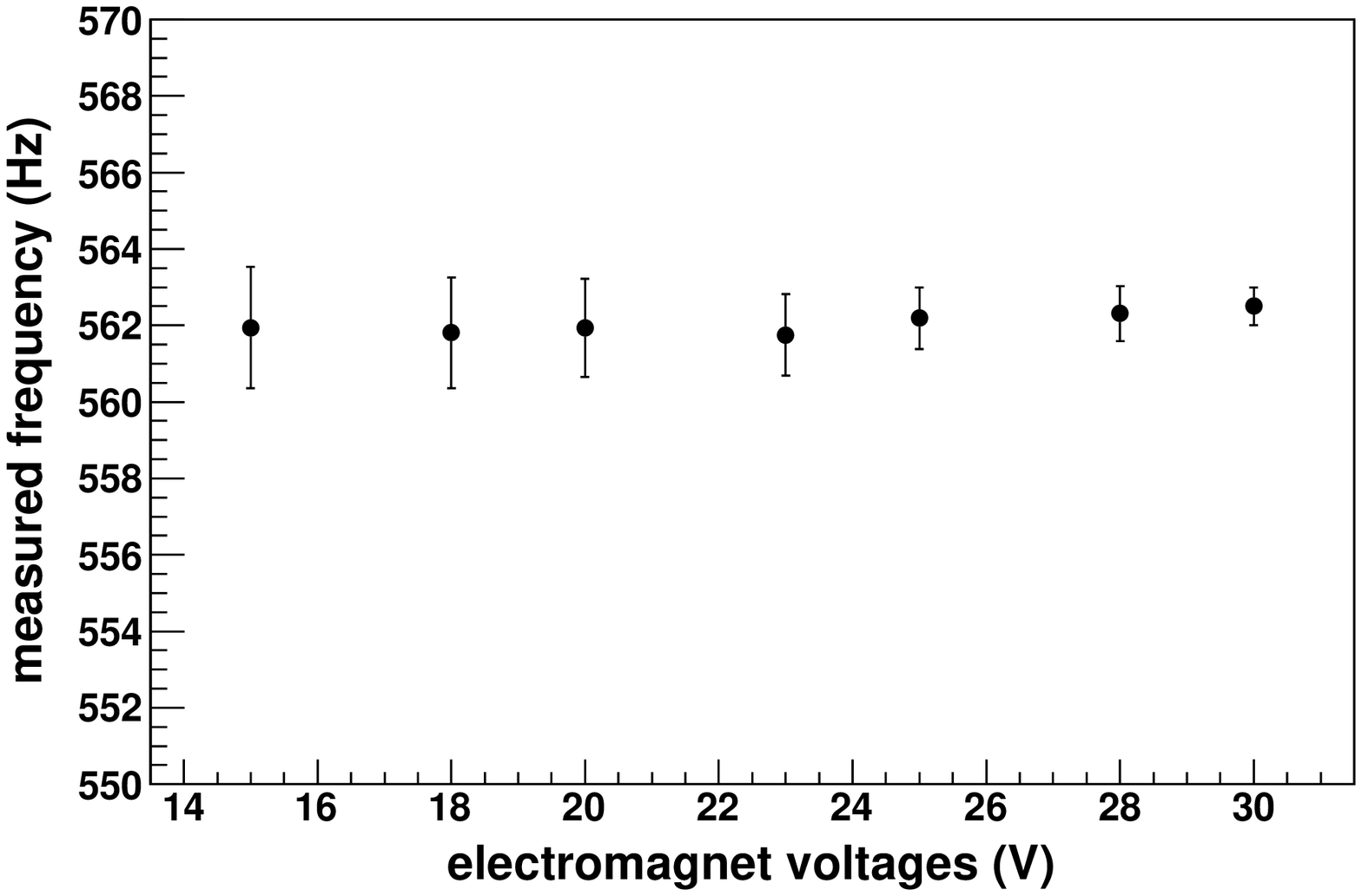}
\figcaption{\label{fig8} The signal amplitude and the average measured frequency v.s. voltages of the electromagnet.
}
\end{center}
\begin{multicols}{2}

\begin{center}
\setlength{\abovecaptionskip}{0.pt}
\setlength{\belowcaptionskip}{-0.pt}
\includegraphics[width=0.47\textwidth]{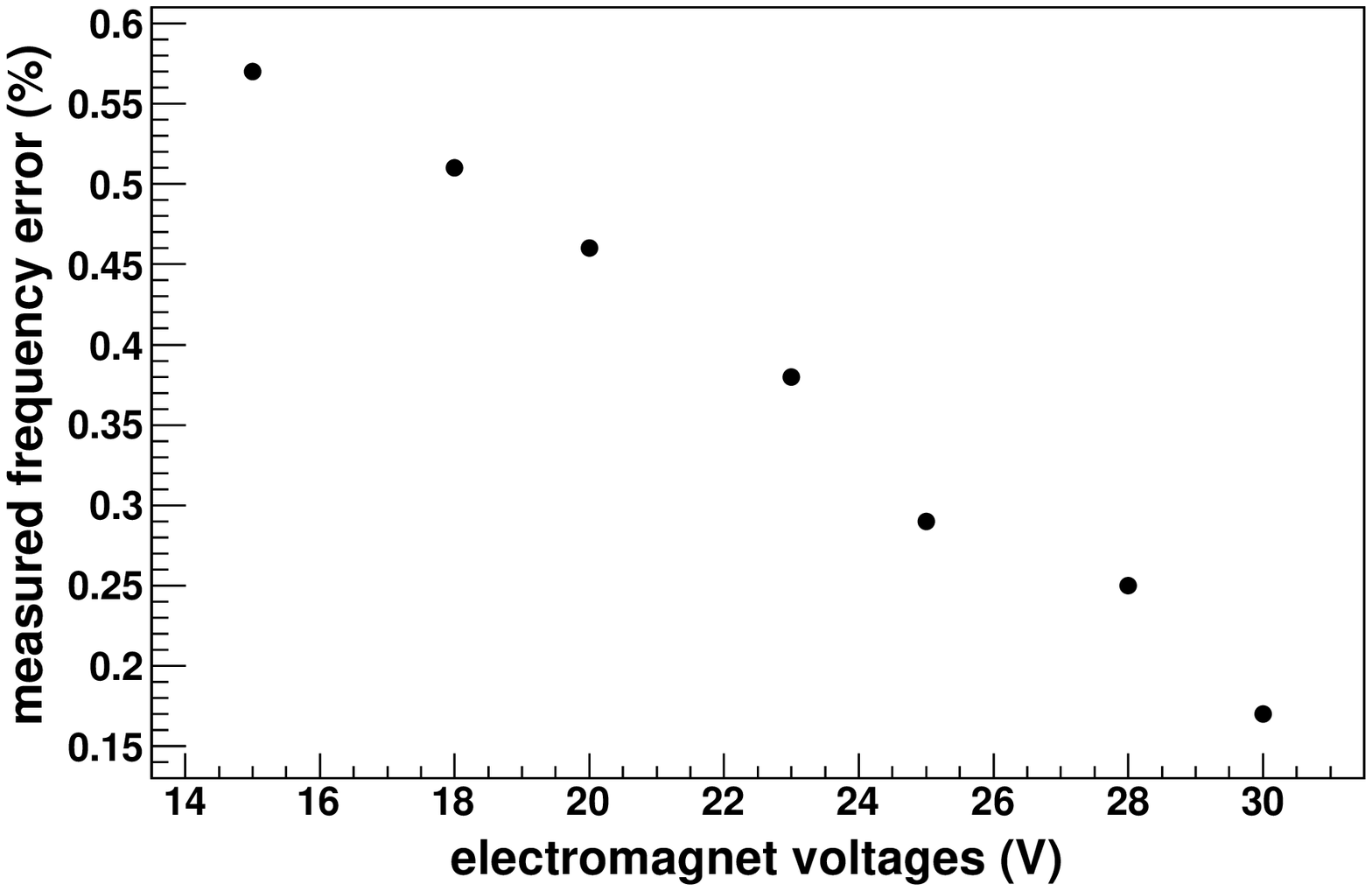}
\figcaption{\label{fig9} The frequency error at different voltages of the electromagnet. }
\end{center}

\begin{center}
\setlength{\abovecaptionskip}{0.pt}
\setlength{\belowcaptionskip}{-0.pt}
\includegraphics[width=0.47\textwidth]{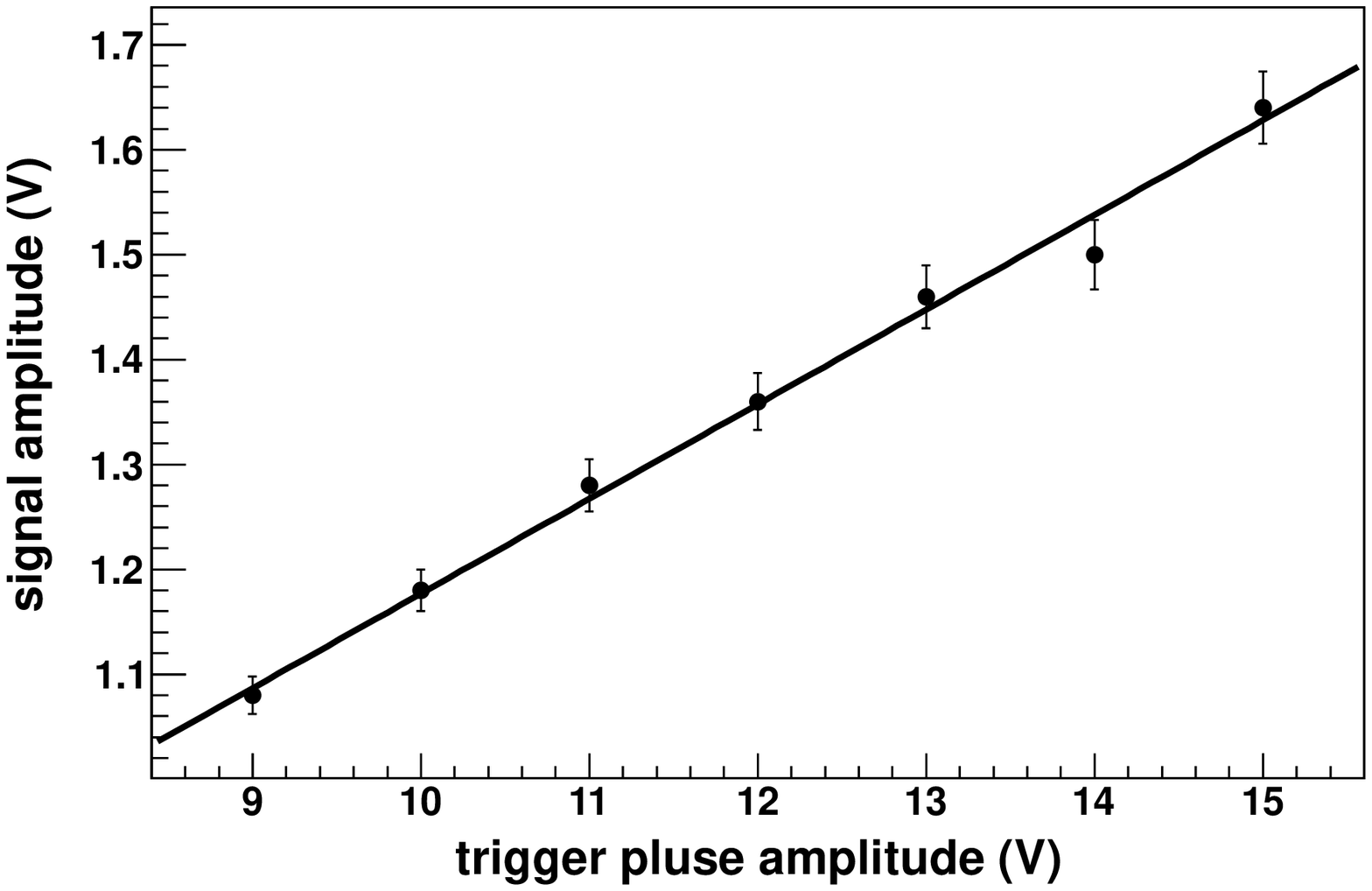}
\figcaption{\label{fig10} The trigger pulse amplitude v.s. the amplitude of the wire signal. }
\end{center}

\subsection{Wire length effect}

For testing the performance the system in measuring different size MWPC, we measured
some 25$\mu$m diameter wires with 50g tension, and the length of the wires is from
250mm to 450mm. In these measurements, the electromagnets' voltage is 30V. Fig.\ref{fig11} is the results, as displayed, the system can get correct frequencies for different length wires. As the results shown, wire's tension of  MWPC with sensitive area from 200mm$\times$200mm to 450mm$\times$450mm can be measured by the system correctly.
\begin{center}
\includegraphics[width=0.47\textwidth]{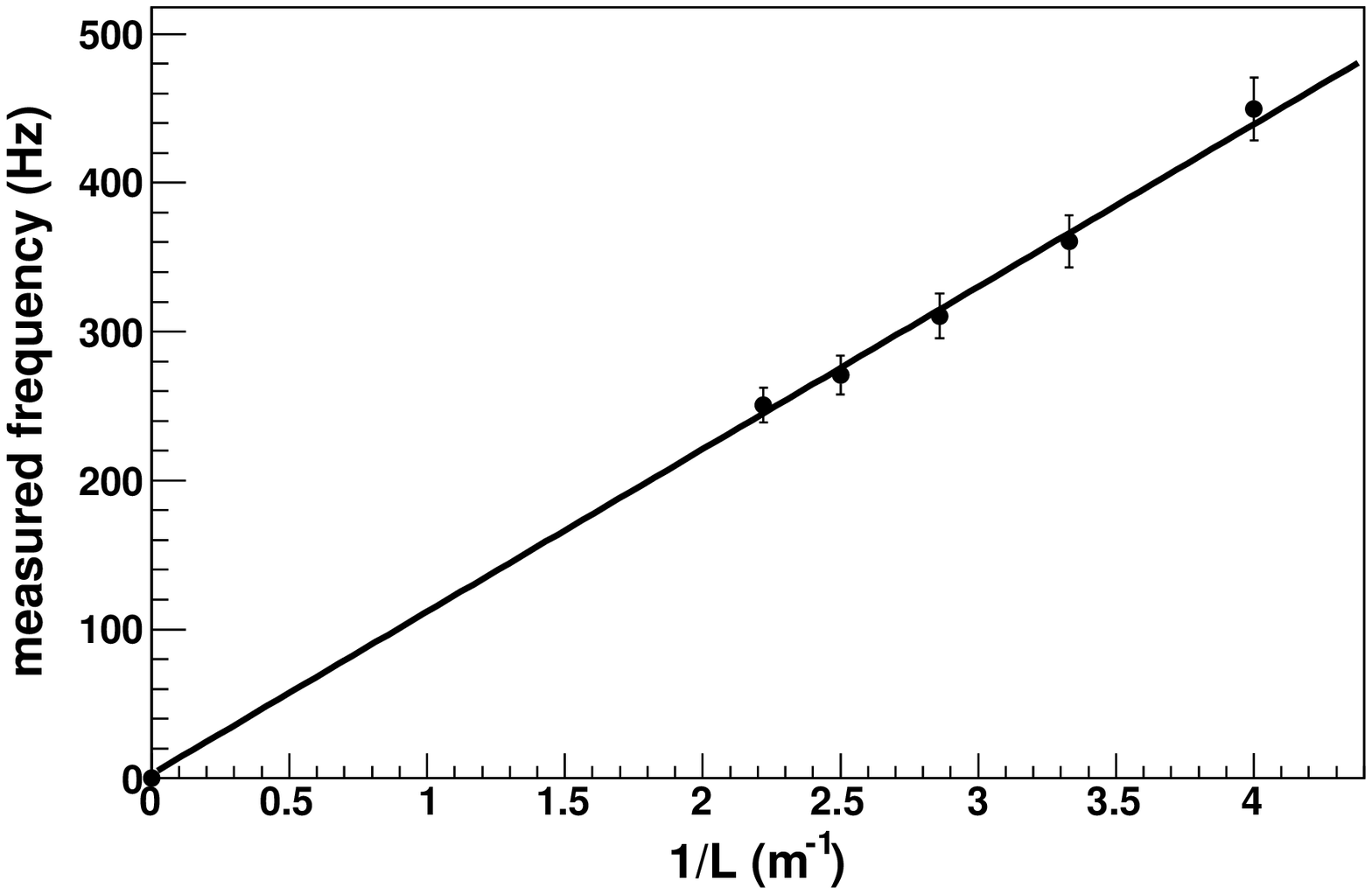}
\figcaption{\label{fig11} The average measured frequency v.s. the wire length's reciprocal.
}
\end{center}

\subsection{The wires' tension of an anode wire plane}

In the MWPC neutron detector of MR in CSNS, many 25$\mu$m, 206mm long wires with 50g tension are welded on a PCB board parallel with 2mm intervals to make an anode wire plane. To make sure the detector have steady and uniform amplification factor, the tension's deviation is controlled to no larger than 10\%. Using the system, we measured the wires' tension. The results are shown in Fig.\ref{fig12} and the standard deviation is about 3.5\%. This mean the wire tension control techniques in the wire plane manufacture is reliable and accurate.

\end{multicols}
\begin{center}
\setlength{\abovecaptionskip}{0.pt}
\setlength{\belowcaptionskip}{-0.pt}
\begin{overpic}[width=0.47\textwidth]{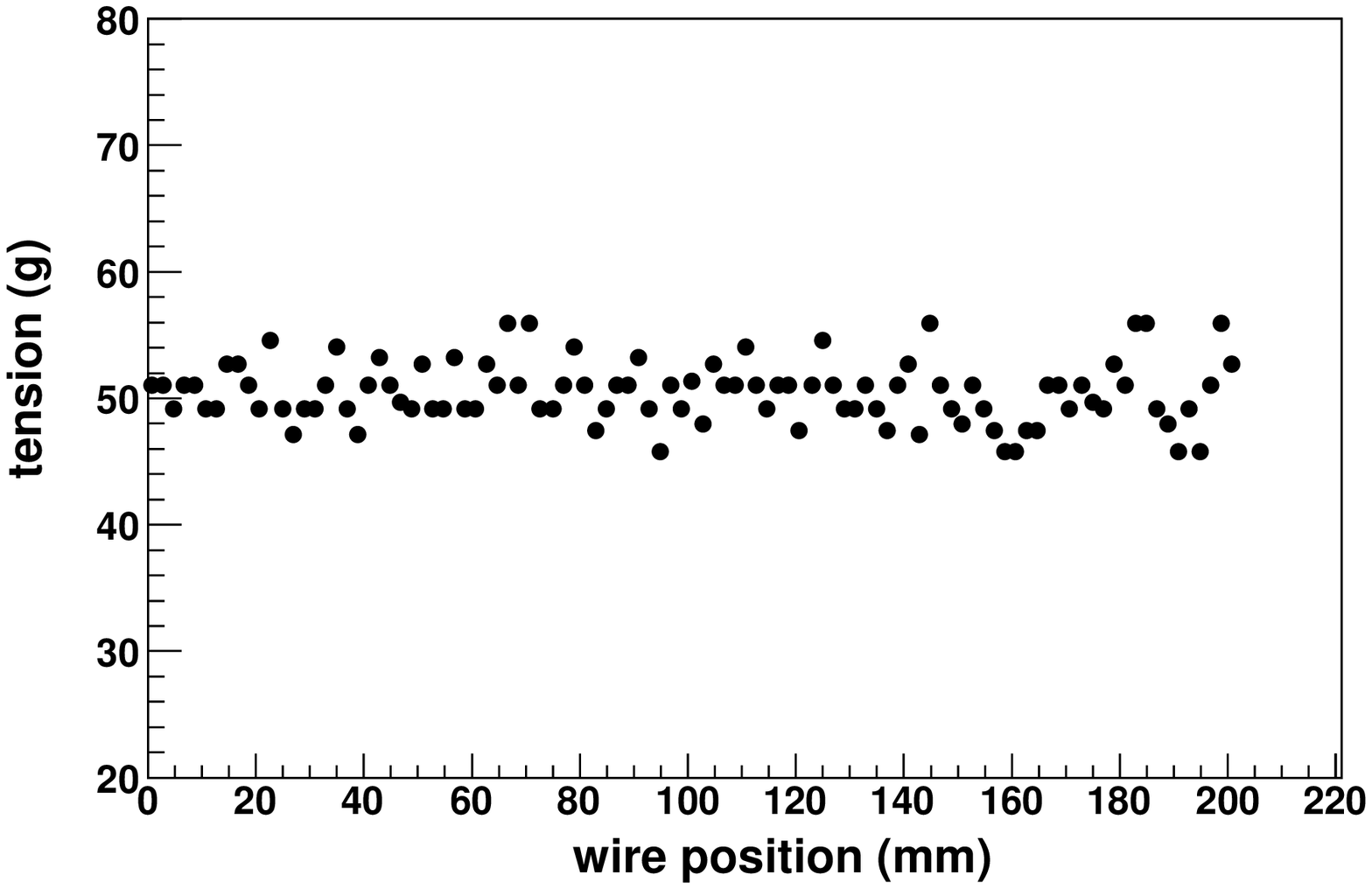}
\put(13, 55){(a)}
\end{overpic}
\begin{overpic}[width=0.47\textwidth]{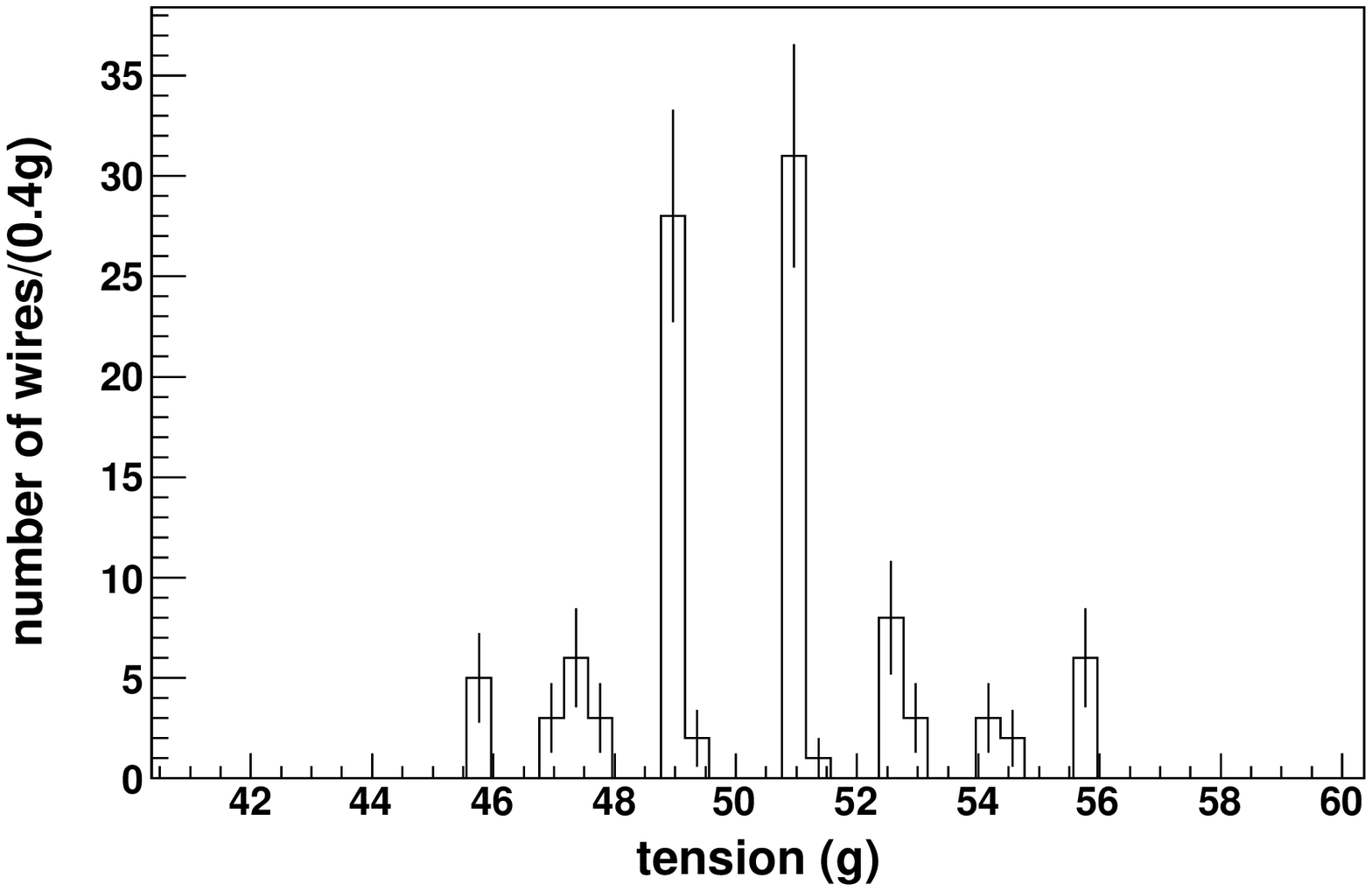}
\put(13, 55){(b)}
\end{overpic}
\figcaption{\label{fig12} The measured wire tensions of an anode wire plane. (a) X axis is the wires' position in millimeter; (b) measured results distribution.}
\end{center}
\begin{multicols}{2}

\section{Discussion}

For the construction of MWPC neutron detectors in CSNS, we developed a portable, accurate and time-saving wire tension measurement system. With a low-voltage power supply and a digital oscilloscope or a personal computer, the system can measure a wire's tension in a few seconds and the error is about 3Hz. Compared with the wire's diameter and tension uncertainties, the measurement error is very small and can be neglected. We also use the system testing all wires' tension in an anode wire pane of a MWPC, the results shown the wire tension control methods in the detector construction is reliable, and they can make the detector work steady in the future.\\

 \acknowledgments{We acknowledge the support from National Natural Science Foundation of China, State Key Laboratory of Particle Detection and Electronics and Key Laboratory of China Academy of Engineering Physics.
 The design of the system consults the paper A\ Device\ for\ Quick\ and\ Reliable\ Measurement\ of\ Wire\ Tension in Princeton/BaBar TNDC-96-39 by Mark R.Convery. We express our appreciation to the author.
 }

\end{multicols}

\vspace{15mm}

\begin{multicols}{2}

\end{multicols}

\vspace{-1mm}
\centerline{\rule{80mm}{0.1pt}}
\vspace{2mm}

\begin{multicols}{2}

\end{multicols}

\clearpage
\end{document}